\documentclass{IEEEcsmag}

\usepackage[colorlinks,urlcolor=blue,linkcolor=blue,citecolor=blue]{hyperref}
\expandafter\def\expandafter\UrlBreaks\expandafter{\UrlBreaks\do\/\do\*\do\-\do\~\do\'\do\"\do\-}
\usepackage{upmath,color}

\usepackage{listings}
\jvol{XX}
\jnum{XX}
\paper{8}
\jmonth{Month}
\jname{Publication Name}
\jtitle{Publication Title}
\pubyear{2021}

\begin{document}

\sptitle{Defining A New Cross Reality: Digital Twins and Mixed Reality Worlds}

\title{MetaGadget: An Accessible Framework for IoT Integration into Commercial Metaverse Platforms}

\author{Ryutaro Kurai}
\affil{Cluster, Inc., Tokyo, 141-0031, Japan and Nara Institute of Science and Technology, Nara, 630-0192, Japan}

\author{Hikari Yanagawa and Yuichi Hiroi}
\affil{Cluster Metaverse Lab, Tokyo, 141-0031, Japan}

\author{Takefumi Hiraki}
\affil{Cluster Metaverse Lab, Tokyo, 141-0031, Japan and University of Tsukuba, Ibaraki, 305-8550, Japan}

\markboth{Defining A New Cross Reality: Digital Twins and Mixed Reality Worlds}{Defining A New Cross Reality: Digital Twins and Mixed Reality Worlds}

\begin{abstract}\looseness-1While the integration of IoT devices in virtual spaces is becoming increasingly common, technical barriers to controlling custom devices in multi-user Virtual Reality (VR) environments remain high, particularly limiting new applications in educational and prototyping settings. We propose MetaGadget, a framework for connecting IoT devices to commercial metaverse platforms that implements device control through HTTP-based event triggers without requiring persistent client connections. Through two workshops focused on smart home control and custom device integration, we explored the potential application of IoT connectivity in multi-user metaverse environments. Participants successfully implemented new interactions unique to the metaverse, such as environmental sensing and remote control systems that support simultaneous operation by multiple users, and reported positive feedback on the ease of system development. We verified that our framework provides a new approach to controlling IoT devices in the metaverse while reducing technical requirements, and provides a foundation for creative practice that connects multi-user VR environments and physical spaces.
\end{abstract}

\maketitle

\chapteri{T}he miniaturization and cost reduction of IoT devices, along with the widespread adoption of cloud services, have accelerated the digitization of smart homes and office environments. At the same time, the concept of digital twins~\cite{Botin-Sanabria2022-zd}, which synchronizes information from physical environments with virtual reality (VR) environments via IoT devices, shows promising applications in industry~\cite{Tao2019-ia} and education~\cite{Sepasgozar2020-ly}. In particular, the implementation of digital twins in multi-user VR environments enables collaborative work where multiple experts and learners can simultaneously analyze and control information from both physical and VR spaces~\cite{Garcia2022-pa}.

However, developers with limited technical knowledge face significant challenges in building such digital twin applications, especially in implementing real-time operation and state synchronization for multiple users. Most existing digital twin systems are implemented as standalone VR environments, requiring advanced technical skills in network programming and database management to build multi-user environments. This technical barrier severely limits the creation of new applications, especially in educational and prototyping environments.

The recent development of metaverse platforms offers a potential solution to these challenges~\cite{Ritterbusch2023-ct}. These platforms abstract complex technical requirements such as multi-user environments and cross-platform compatibility from VR-HMDs, PCs and smartphones, making them accessible to general users. However, existing commersial metaverse platforms primarily focus on interactions only within VR spaces~\cite{Slater2022-mr}, with very limited functionality for connecting to physical IoT devices. While some platforms such as VRChat~\cite{vrchat2023} support device control through Open-Sound Control (OSC) and serial communication~\cite{Wright1997-ex}, they lack well-maintained frameworks and still require significant technical expertise.

To address these challenges, we propose MetaGadget, an IoT device control framework for commercial metaverse platforms. MetaGadget uses HTTP-based event triggers to provide loose synchronization between physical devices and virtual spaces, maintaining functionality even when users are offline. This framework eliminates the need to maintain persistent server connections or adhere to specific communication protocols, enabling developers and educators without programming expertise to create cross-platform IoT device control applications in multi-user environments.

To evaluate the effectiveness of our proposed framework, we conduct two workshops targeting users with different technical skill levels. The first one-day workshop is aimed at users ranging from novice to intermediate programmers, and focuses on the rapid development of applications that connect metaverse platforms to SwitchBot~\cite{switchbot2023}, a commercial IoT device. The second workshop, lasting approximately one month, is designed for advanced users with experience in VR and IoT development, and includes the design and implementation of custom devices using Raspberry Pi. Through these workshops, we will examine the efficiency of developing digital twin applications using MetaGadget in metaverse environments, while exploring new possibilities for physical-virtual integration in multi-user access environments.

We have previously demonstrated the basic concept of IoT-metaverse integration~\cite{Kurai2024-mr}, which showed the feasibility of controlling physical devices through metaverse platforms. 
While our previous work focused on demonstrating the basic concept with custom-built devices, this paper presents a mature framework that supports commercial IoT devices like SwitchBot and provides systematic development support. 
Furthermore, we conduct comprehensive empirical validation through two workshops with users of varying technical expertise, whereas the previous work was limited to technical demonstration.


\section{METAGADGET FRAMEWORK}
\begin{figure*}[t]
    \centering
    \includegraphics[width=1\linewidth]{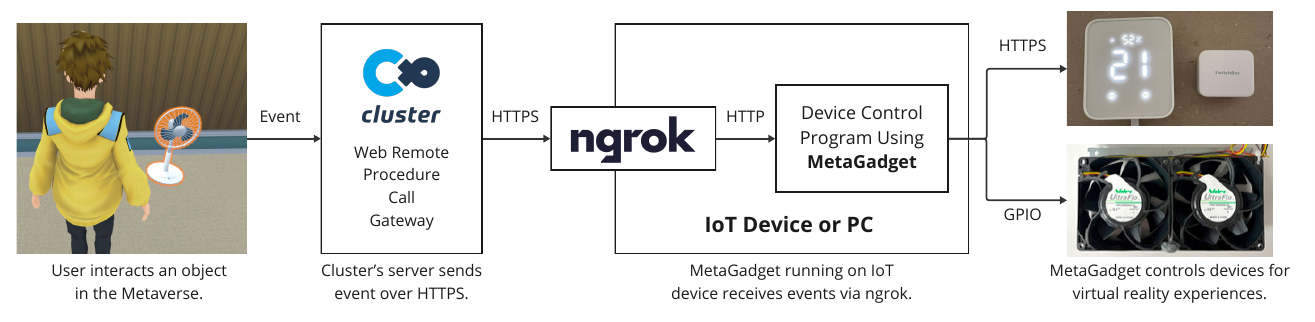}
    \caption{MetaGadget System Overview}
    \label{fig:system_overview}
\end{figure*}

Our MetaGadget framework converts and transmits user input from metaverse platforms into control signals for external IoT devices. A key feature of this framework is its adoption of HTTP-based asynchronous communication, making it implementable on any metaverse platform capable of sending HTTP POST requests to external server URLs.

In this research, we implemented the framework on Cluster~\cite{cluster}. Cluster is a multi-platform metaverse platform that runs on standalone Windows/MacOS applications, VR headset environments such as Meta Horizon OS, and smartphone platforms such as iOS and Android. Also, Cluster supports its proprietary script, Cluster Script, a JavaScript-based programming language. By attaching this script to objects in VR spaces, developers can define how objects respond to user interactions. In particular, Cluster Script includes \verb|callExternal()| function for executing HTTP POST requests to external servers, making it particularly compatible with our MetaGadget framework.

\subsection{MetaGadget FrameWork}
Figure~\ref{fig:system_overview} shows the overview of our MetaGadget system.
MetaGadget is distributed as a Python library via the PyPI package\footnote{\url{https://pypi.org/project/metagpt/}} and implements a server that receives POST requests from the Web Remote Procedure Call Gateway. It runs on any operating system that supports Python, including Windows, Linux, and MacOS.

Users can easily implement MetaGadget Server using the following code:
\begin{lstlisting}[language=Python]
from metagadget import MetaGadget
app = MetaGadget()
app.run()
\end{lstlisting}

When \verb|app.run()| is executed, it launches Werkzeug\footnote{\url{https://werkzeug.palletsprojects.com/en/stable/}}, a utility library that processes HTTP requests and responses as manageable objects while providing HTTP server functionality. At the same time, it launches ngrok, a tunneling service for exposing the server to the Internet. This allows the server to run on computers without public HTTP ports and generates a unique URL.

Users can process data POSTed to the server and return responses in JSON format or string using the following code:
\begin{lstlisting}[language=Python]
@app.receive
def handle(data): 
    # Server receives data
    result = process_input(data)
    # Prepare response in JSON format
    ret = prepare_response(result)
    return ret
\end{lstlisting}

\subsection{Metaverse Platform Integration}
The \verb|callExternal()| function in Cluster Scripts executes in the client's local environment. This function POSTs its string argument as part of a JSON structure to a preset URL via the Web Remote Procedure Call Gateway. By setting this URL to the one issued by MetaGadget and linking the function to VR scene objects, control signals can be sent to the server when specific events occur.

\subsection{Physical Device Control}
Physical devices connected to the MetaGadget server can be controlled based on received control signals. For example, electronic components connected through GPIO (General Purpose Input/Output) pins on a Raspberry Pi can be controlled and sensor values can be read. Here is an implementation example of LED ON/OFF control:
\begin{lstlisting}[language=Python]
@app.receive
def handle(data):
    if data == "on":
        GPIO.output(LED_PIN, 1)
    else:
        GPIO.output(LED_PIN, 0)
\end{lstlisting}

Our framework can also control devices managed by external services by calling their APIs. For example, SwitchBot provides a variety of IoT devices with flexible web APIs. These include light bulbs, electric fans, power strips that can be controlled via API, and robots that can operate physical switches, all of which can be controlled via web APIs. 
MetaGadget also provides sample code to control these SwitchBot devices as:
\begin{lstlisting}[language=Python]
@app.receive
def handle(data):
    # Parse request for SwitchBot control
    sh_req = SmartHomeRequest.model_validate_json(data)
    try:
        # Retrieve and Execute requested function
        func = getattr(switchbot_client, sh_req.function_name)
        ret = func(*sh_req.args, **sh_req.kwargs)
        # Convert response to JSON
        if isinstance(ret, BaseModel):
            ret = ret.model_dump_json()
        return ret
    except Exception: 
        ...
\end{lstlisting}

\section{APPLICATION EXAMPLES}
\begin{figure}[b]
    \centering
    \includegraphics[width=1\linewidth]{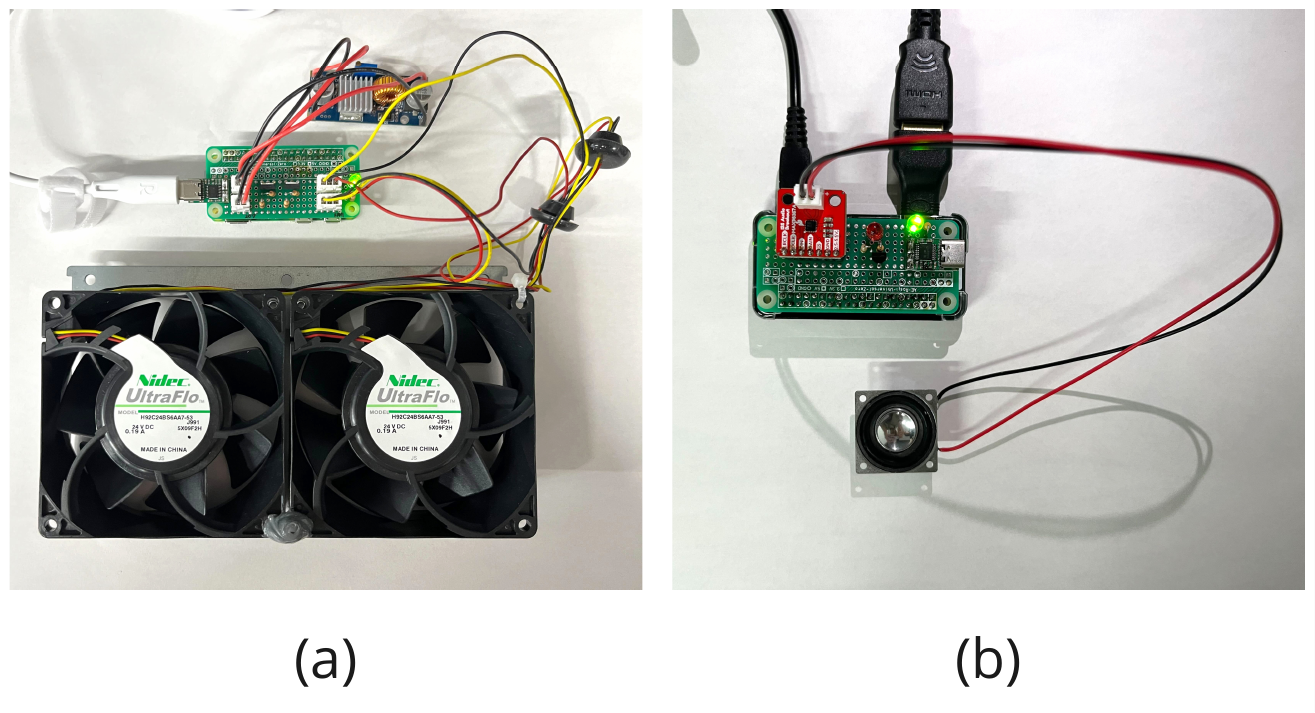}
    \caption{(a)An electronic fan device controlled from metaverse space. (b) A speaker device built as a doorbell. Both devices are connected to control units running MetaGadget.}
    \label{fig:phisical_devices}
\end{figure}

To demonstrate the practical applications of the MetaGadget framework, we have developed two implementation examples that integrate physical devices using Raspberry Pi with the Cluster platform, as shown in Figure \ref{fig:phisical_devices}. These implementations take advantage of Cluster's multi-platform compatibility, allowing control from both laptops and smartphones. We created a VR space containing interactive VR objects corresponding to each physical device. Users can access this VR space through Cluster to experience devices that bridge the physical and virtual environments.

\paragraph{Fan} We also implemented a virtual fan that is synchronized with a physical fan. This system controls the rotation of two PC fans connected to a Raspberry Pi. User interaction with the virtual fan object in the VR space is immediately reflected in the rotation and stopping of the physical fan through the MetaGadget server. All users in the VR room can control this virtual fan, allowing for shared device control by multiple users.

\paragraph{Doorbell} We implemented a room-entry notification system that demonstrates event-triggered interaction between VR spaces and physical devices. When a user enters the house in the VR space, the Cluster Scripts attached on the floor detect this entry event and send a signal to the MetaGadget server, which then plays a chime sound through a speaker connected to the Raspberry Pi. A key feature of this system is that it functions independently of the device owner's login status, demonstrating the framework's ability to control devices through event-triggered interactions without requiring persistent connections.

\section{WORKSHOP AND FEEDBACKS}
To evaluate the effectiveness of the proposed MetaGadget framework, we conducted two workshops with distinct objectives. The evaluation focused on two main research questions:
\begin{enumerate}
    \item Can non-technical users build systems to control IoT devices from metaverse environments using MetaGadget?
    \item Can users create new forms of interaction in digital twin applications using MetaGadget, taking advantage of the real-time multi-user operations that are a key feature of metaverse platforms?
\end{enumerate}

\section{WORKSHOP FOR NON-TECHNICAL USERS}
The first workshop was a one-day event where participants with different levels of development experience, from novice to advanced, developed applications that connected SwitchBot devices to Cluster using MetaGadget. Then we conducted the questionnaires and semi-structured interviews to evaluate the usability, potential application and improvement. 


\subsection{Workshop Setups}
Participants developed applications that connected smart home appliances with the metaverse using MetaGadget-enabled SwitchBot devices. The framework implementation enabled physical device control through HTTP communication, where click events on items placed in Cluster were transmitted to the SwitchBot hub. By abstracting these event control and external communication mechanisms within the framework, we created an environment where even novice programmers could implement device control.


We provided the following SwichBot devices as the development environment:
\begin{itemize}
    \item Control Infrastructure: SwitchBot Hub (2 units)
    \item Sensors: Smart cameras (2 units), Motion sensors (4 units), a CO2/Temperature/Humidity sensor
    \item Actuators: LED Strip Lights, Smart Bulbs (4 units), Light Sockets (4 units)
    \item Interfaces: Smart Plugs (4 units), SwitchBots (4 units)
    \item Environmental Control: One Humidifier, Remote Buttons (2 units), Circulators (2 units)
\end{itemize}


To simplify application development, we used Cluster's World Craft feature, which allows direct editing of 3D space within the metaverse environment by placing craft items (3D objects) without the need for external development tools such as Unity. We prepared and distributed craft items with pre-integrated ClusterScripts controlling SwitchBot to allow novice programmers to participate in application design. For intermediate and advanced participants, we provided additional flexibility by allowing custom craft item uploads via Unity and ClusterScript coding.


\subsection{Participant Demographics}

The workshop included 14 participants (mean 30.0~$\pm$~6.7 years, 7 male and 7 female). Participants self-reported their technical skills as: 
\begin{itemize}
\item Advanced (capable of independent development): 1 participant (P8)
\item Intermediate (capable with assistance): 7 participants (P1, P2, P3, P6, P7, P11, P13)
\item Novice (limited to basic use of crafting items): 6 participants (P4, P5, P9, P10, P12, P14)
\end{itemize}
The distribution of detailed technical skills is as follows:
\begin{itemize}
    \item Unity Development Experience: 1 participant (P8) had 3 more years experience, 4 participants (P1, P2, P7) had worked with Unity for 1-3 years, 4 patricipants (P6, P10, P13, P14) had less than 1 year, and 5 participants had no prior experience.
    \item VR Platform Experience: 4 participants (P1, P8, P10, P11) had previously created VR worlds. 5 participants (P2, P3, P5, P7, P12) had only visited VR environments without content creation experience, while the remaining 5 had no prior exposure to VR platforms.
    \item IoT Device Development Experience: 5 participants (P6, P7, P10, P13, P14) reported having previous experience, while 9 participants had never worked with IoT devices.
\end{itemize}






\subsection{Workshop Structure}
The one-day workshop offered participants a compensation of 10,000 JPY. The program consisted of three sections, with clearly indicated difficulty levels to accommodate different skill levels:

\begin{enumerate}
\item Orientation (90 minutes): All participants received a comprehensive overview of Cluster and MetaGadget. This was followed by beginner-level training focused on basic craft item placement and SwitchBot control. Finally, intermediate and advanced participants engaged in custom model creation using Unity and the Cluster API.
\item Application Development (180 minutes): Participants worked individually or in groups to design and implement IoT device and metaverse integration applications suitable for multi-user experiences.
\item Presentation and Feedback (30 minutes): Participants presented the applications they had developed and answered the questionnaire described below.
\end{enumerate}

After the workshop, participants completed a questionnaire consisting of 5-point scale items and open-ended questions. In addition to the System Usability Scale (SUS)~\cite{Brooke1996-kw}, three specific criteria were included:
\begin{enumerate}
    \item[Q1.] I could control physical devices from virtual space using this framework.
    \item[Q2.] I found it easy to create synchronized multi-user systems with this framework.
    \item[Q3.] I was able to implement the applications I had in mind.
\end{enumerate}







\subsection{Framework Evaluation}
We evaluate the MetaGadget framework, examining both its technical capabilities and user experience aspects. The evaluation focuses on three key areas: core functionality, technical scope, and overall usability of the development environment.

\paragraph{Core Function Rating}
The system received positive scores for both its ability to control devices (Q1) and its ease of deployment in multi-user environments (Q2). For device control, the overall average score was 4.57 out of 5 (novice: 4.00, intermediate and advanced: 4.89). Similarly, the overall score for multi-user implementation was 4.43 (novice: 4.00, intermediate and advanced: 4.67).

These quantitative results were supported by qualitative feedback: novices appreciated the accessibility of the system: "It was easy because I could just copy and paste scripts from the template world" (P14), "The basic operating code was well explained and understandable" (P4). Intermediate users also recognized the benefits and compared it favorably to traditional methods: "Compared to my usual approach of using MQTT with Raspberry Pi, this requires much less technical expertise." (P7)



\paragraph{Implementation Completeness and Barriers}
Implementation Completeness (Q3) received a lower average score of 3.57, with a notable gap between novices (2.60) and more experienced users (4.11). This disparity reflects P3's comment that "understanding and effectively using the code becomes a challenge when attempting more complex implementations. While basic functionality was accessible, advanced implementations still present technical barriers.



\paragraph{Development Environment Usability}
The System Usability Scale (SUS) yielded an average score of 54.6~$\pm$~16.2, placing the system in the "Low Marginal" category according to the Acceptability Ranges. Note, however, that this low rating reflects not only the usability of the MetaGadget framework itself, but also the development experience of the platform as a whole. 

In fact, when participants were asked which stages of development took the most time, many cited platform-related challenges. Users who had no previous experience with virtual world creation often mentioned that "world creation was time consuming" (P2, P3). P1 noted that "distinguishing between cluster-specific components, JavaScript elements, and the proposed framework components proved challenging," suggesting that platform complexity affects the development experience.

Specific to the MetaGadget framework, participants identified several areas for improvement. P2 and P8 recommended strengthening the error handling capabilities. In addition, several participants (P3, P4, P14) requested more code examples and comprehensive API documentation for SwitchBot integration. These emerged as important future improvements for novice developers learning the system.





\subsection{Emerging Interaction Patterns and Applications}
Our workshop results show that the MetaGadget framework goes beyond traditional digital twin concepts; in particular, P7 pointed out that our framework has the potential to create new value by connecting multiple physical spaces with multiple VR spaces in various configurations. Based on this insight, we analyze the participants' implemented and proposed applications from two perspectives: application domains and spatial connection patterns.


\paragraph{Social Support through Multi-Space Connections}
\begin{figure*}[t]
    \centering
    \includegraphics[width=1\linewidth]{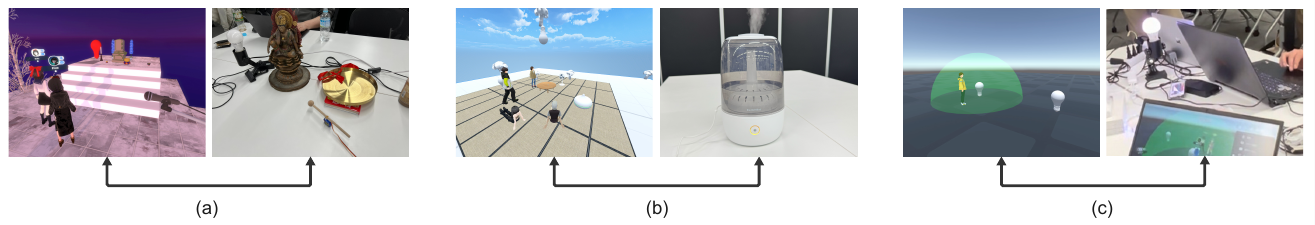}
    \caption{Some examples of systems connected to the metaverse created by the participants. (a) A grave in the metaverse space is connected to a real-life religious object, allowing people to pray with a sense of realism. (b) From the metaverse space, where it is possible to communicate with the elderly, it is possible to operate a real-life home appliances in response to the needs of the elderly. (c) When people gather in the green dome in the metaverse space, the lights in the real world become brighter, indicating the number of people in the real world.}
    \label{fig:connected_system}
\end{figure*}
In social support applications, we observed value creation that scaled with the complexity of the connection pattern. The simplest implementation was P10's remote grave visitation system (Fig.~\ref{fig:connected_system}, a). Although this system used a simple one-to-one connection between a metaverse space and physical space, it revealed the potential of our framework as "a sense of spiritual security in being able to feel a physical connection to the grave even when we are too busy to visit the grave" (P10).

P10 evolved this idea further, proposing a "remote toy control system for entertaining children" and introducing cooperative control with family and friends. This emphasized the benefit of being able to keep an eye on children without worrying about the physical state of the house (e.g., how messy it is).

One group (P6, P7, P11, P13, P14) implemented a "home security and remote monitoring system" (Fig.~\ref{fig:connected_system},  b). They emphasized the advantage of being able to "actually see the person operating the physical device in VR, which gives the elderly a sense of security that someone is watching over them". This is similar to an actual elder care facility and demonstrates the effectiveness of a many-to-many connection pattern, where multiple experts from different metaverse spaces collaborate to manage the environment of multiple beneficiaries.

This many-to-many connection pattern also shows promise in education; P13 proposed a collaborative environmental learning system where multiple educators jointly control multiple learning environments, suggesting new possibilities in educational support.

\paragraph{Cross-Space Presence Representation and Environmental Sharing}
Participants also implemented and proposed various applications for visualizing presence and sharing environments. Proposed applications such as "Displaying visitor notifications on a smartwatch in VR Space" (P1) and "Integrating voice control with smart speakers" (P12) take advantage of MetaGadget's ability to control physical devices without logging in to the platform, and show that even simple connections can enable effective and natural presence sharing.

In a more complex example, P1 and P8 separately implemented the idea that the brightness of a physical light bulb increases as more people gather in the metaverse space~(Fig. \ref{fig:connected_system}, c). This succeeded in expressing the collective existence of participants from multiple metaverse spaces through the state of a single physical device. P12 focused on the potential of this concept in the art scene, suggesting the possibility of creating cultural value through the visualization of existence.

The creation of a sense of unity through the sharing of multimodal experiences in multiple physical spaces was also discussed. P4 suggested that "metaverse space and physical lighting, scent, temperature, and humidity be synchronized", and P5 suggested specific applications such as "adjusting the temperature of rooms in all participants' physical space depending on location, such as areas with magma or ice".

\paragraph{Entertainment Applications of Multi-Space Interaction}
The entertainment sector highlighted particularly interesting applications of many-to-many connections. Inspired by YouTube Live's SuperChat system, P7 proposed a system where "viewers can control IoT devices in the streamer's space based on SuperChat contributions. This creates a new form of interaction where multiple viewers from different metaverse spaces influence a single performer's environment.

P14 extended this concept further by proposing a "fan service" that "creates a sense of specialness by allowing multiple performers to select individual participants and trigger actions on their physical devices," presenting an advanced example of a many-to-many connection between viewers and performers.

P7 also envisioned of a system in which home appliances would respond to the atmosphere of a group conversation. This suggests the possibility of integrating AI to naturally reflect the status of multiple participants in multiple physical spaces.



\section{WORKSHOP WITH CUSTOME DEVICE DEVELOPMENT}
The second workshop was conducted over a month as part of a university hands-on course. In this workshop, students with VR development experience designed custom devices connected to Raspberry Pi and developed applications that integrated these devices into the metaverse using MetaGadget. The workshop concluded with participant interviews to evaluate the framework.


\subsection{Workshop Setups}

Participants designed and developed experiences connecting custom hardware devices attached to Raspberry Pi with the metaverse using MetaGadget. The MetaGadget framework abstracts event control and external communication mechanisms within the metaverse space, enabling application development without requiring extensive hardware expertise.
Each participant received the following development hardware:

\begin{itemize}
\item Single-board Computer: Raspberry Pi Zero 2 W including Micro SD card with Raspbian OS and USB MicroB cable.
\item Electronics Prototyping Kit: DaVinci Kit for Raspberry Pi (SunFounder), which included essential prototyping components such as breadboards, LEDs, speakers, and motors.
\end{itemize}

Participants used their personal computers, with laptops (Mac, Windows) available upon request.
They connected to their Raspberry Pi via SSH over USB, accessing the internet through their PC's network sharing capabilities.
All participants had prior experience with Cluster Creator Kit (CCK), Cluster's Unity-based software development kit, which they used for application development.





\subsection{Participant Demographics}
The workshop included five male undergraduate students from the University of Electro-Communications in Japan, comprising four second-year and one first-year students.
A single faculty member conducted the lectures.
All participants had completed a six-month hands-on course on metaverse world creation using Unity and CCK prior to the workshop, with each having created their original worlds.
They also had hardware prototyping experience from other practical courses.


\subsection{Workshop Structure}
The workshop was integrated into the ``Information Engineering Laboratory'' course at the University of Electro-Communications, spanning approximately one month.
Each 90-minute session included classroom orientation and question-answering periods, with practical work assigned as homework. 
A Slack channel was established for instructor support outside of class hours.
The weekly program was structured as follows:
\paragraph{Week 1: Orientation}
\begin{itemize}
\item Hardware distribution and usage instruction
\item Development environment setup connecting PC and Raspberry Pi
\item MetaGadget overview and tutorial completion
\end{itemize}
\paragraph{Weeks 2--4: Progress Reports and Development}
\begin{itemize}
\item Documentation-based progress reporting
\item Development work
\item Technical support and problem-solving
\end{itemize}
\paragraph{Week 5: Final Presentations}
\begin{itemize}
\item Presentation of developed metaverse experiences
\item In-class demonstrations
\end{itemize}




 
\subsection{Interaction and Applications with Custome Hardware Devices}
\begin{figure*}[t]
    \centering
    \includegraphics[width=1\linewidth]{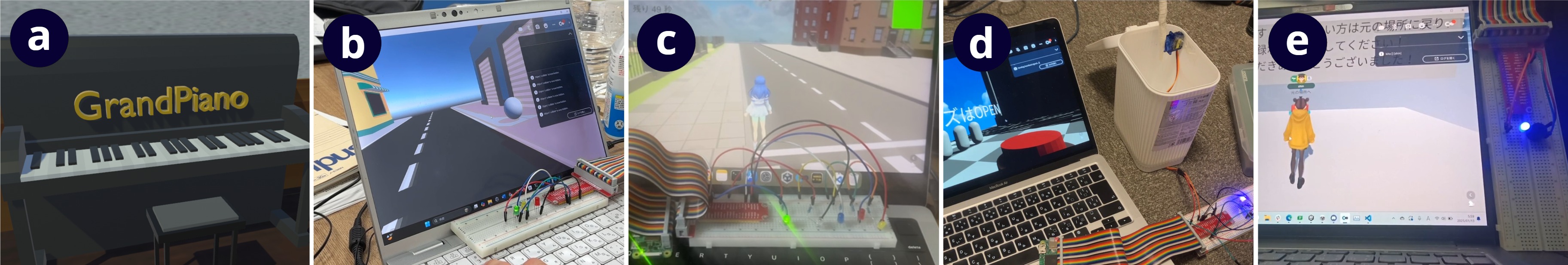}
    \caption{Applications developed in the workshop. (a) A virtual piano system with frequency-controlled sound output (P1). Each key press triggers a corresponding audio frequency ranging from 110Hz to 1396.91Hz. (b) A traffic signal system with integrated LED and servo motor control (P2). The system enables pedestrian crossings synchronized with virtual space interactions. (c) A color-based game interface that translates virtual color commands into physical LED outputs (P3), extending the participant's previous metaverse world. (d) An interactive waste management system (P4) that facilitates collaboration between physical and virtual space users through synchronized bin operations. (e) A game enhancement system featuring audio feedback mechanisms (P5), building upon an existing game environment to provide sound-based player interaction cues.}
    \label{fig:ws2_apps}
\end{figure*}

Five participants developed the following applications during the workshop:

\begin{itemize}
\item Virtual piano system (P1)
\item Traffic light control system (P2)
\item Game experience with LED visual effects (P3)
\item Interactive trash bin system (P4)
\item Game experience with sound feedback (P5)
\end{itemize}

Figure~\ref{fig:ws2_apps} illustrates the implementation details of each application, demonstrating how participants integrated physical devices with virtual interactions.

The virtual piano system (P1) featured precise frequency control for musical note reproduction. The system successfully reproduced piano-like playing experiences by outputting corresponding frequencies for each key press, covering a broad frequency range from 110 Hz to 1396.91 Hz with precise pitch control.

The traffic light system (P2) reflected virtual space control inputs in physical traffic light operation using Raspberry Pi. The system implemented simultaneous LED and servo motor control to simulate safe pedestrian crossing, faithfully reproducing real traffic system logic including pedestrian crossing permissions based on signal states.

P3 expanded their previously developed metaverse world by implementing a game with LED visual effects. The system enhanced player interaction by representing virtual space color instructions through physical LED illumination, successfully increasing immersion in the existing world through the addition of physical feedback mechanisms.

The interactive trash bin system (P4) featured a unique design requiring cooperation between users in physical and virtual spaces. The system achieved stable operation through state transition management and enabled cross-space communication through tactile interaction implementation. Future expansion plans include smart home feature integration.

P5 extended their previously developed game by adding sound feedback functionality. The system successfully enhanced the gaming experience through features such as game-over sound effects, achieving integrated operation with the existing system and increasing game immersion through audio feedback.

These applications can be categorized into three main types:
\begin{enumerate}
\item Visual Feedback Systems: they utilize light and motion as primary forms of expression. These systems promote intuitive understanding by translating virtual space operations and states into physical light and movement patterns.
\item Audio Interaction Systems: they employ sound as their primary output medium. These implementations range from accurate frequency-controlled instrument reproduction to game experience enhancement through sound effects.
\item Physical Environment Control Systems: they focus on manipulating physical devices. These systems aim to create tightly integrated connections between virtual and physical reality.
\end{enumerate}

Through analysis of these applications, we confirmed that applications developed through custom device design achieve more flexible interactions between virtual and physical spaces. Beyond simple device on/off operations, these implementations demonstrate efforts to achieve tighter integration through features such as servo motor control.

\subsection{Feedback from Participants}
Analysis of the participant evaluations and feedback revealed three main areas of evaluation:

First, regarding the Cluster/MetaGadget development environment, participants provided feedback on both technical challenges and learning opportunities. P4 specifically reported challenges with "error handling and response format optimization in external server communication from Cluster". This process led to insights into "the importance of communication protocol mechanisms and formatting", confirming the value of the framework as a learning platform.

Second, participants provided positive feedback on the hardware integration development environment, particularly with respect to the Raspberry Pi implementation. While P1 faced challenges in implementing frequency control for audio output, they successfully discovered appropriate implementation methods. P2 reported gaining a deeper understanding of SSH connections and Linux commands, despite having to reinstall the Raspberry Pi OS several times.

Third, in terms of overall development process learning outcomes, several participants reported gaining skills in "understanding and using existing libraries" and "developing the ability to solve problems through independent modification and customization". This feedback suggests that the framework served not only as a development environment, but also as an effective hands-on learning platform.



\section{CONCLUSION}
In this paper, we present MetaGadget, a framework that bridges the gap between IoT devices and metaverse platforms. Our research goes beyond traditional digital twin concepts by enabling many-to-many connections between physical and virtual spaces.

The results of the workshop demonstrated two key findings. First, MetaGadget successfully lowered the technical barriers to IoT device control in multi-user environments, as evidenced by novice users implementing basic device control functions with minimal programming knowledge. Second, the framework enabled experienced developers to create sophisticated applications that demonstrate meaningful interactions between physical and virtual spaces.
For future work, the workshops also revealed challenges that require further investigation, including development support, real-time performance beyond HTTP-based connections, and cross-platform compatibility.

MetaGadget represents a first step toward democratizing IoT-metaverse integration. As metaverse platforms continue to evolve, frameworks that enable the accessible development of physical-virtual interactions will become increasingly important for realizing the full potential of spatial computing.

\vspace*{-8pt}

\section{ACKNOWLEDGMENTS}
This work was partially supported by JST ASPIRE Grant Number JPMJAP2327.


\def\refname{REFERENCES}
\bibliographystyle{IEEEtran}
\bibliography{references}

\begin{IEEEbiography}{Ryutaro Kurai}{\,} 
 is an engineering manager at Cluster, Inc. and a Ph.D. student at Nara Institute of Science and Technology (NAIST), Japan. His current research interests include metaverse, human-computer interaction, and large language models. He received his M.S. degree in computer science from Hokkaido University. He is a Student Member of the IEEE Computer Society. Contact him at r.kurai@cluster.mu.
\end{IEEEbiography}

\begin{IEEEbiography}{Hikari Yanagawa}{\,}
 is a research engineer at Cluster Metaverse Lab, Japan. Their current research interests include reinforcement learning, neural radiance fields, and real-time deep-fake technology. They received their Advanced Diploma in 3D Computer Graphics from HAL Nagoya. Contact them at h.yanagawa@cluster.mu.
\end{IEEEbiography}

\begin{IEEEbiography}{Yuichi Hiroi}{\,} is a senior research scientist at Cluster Metaverse Lab, Japan. His current research interests include augmented reality, near-eye displays and vision augmentation. He received his Ph.D. degree in computer science from Tokyo Institute of Technology. He is a Member of the IEEE Computer Society. Contact him at y.hiroi@cluster.mu.
\end{IEEEbiography}

\begin{IEEEbiography}{Takefumi Hiraki} {\,} is a senior research scientist at Cluster Metaverse Lab and an 
assistant professor at University of Tsukuba, Japan. His current research interests include augmented reality, haptic interfaces, and soft robotics. He received his Ph.D. degree in engineering from the University of Tokyo. He is a Member of the IEEE Computer Society. Contact him at t.hiraki@cluster.mu.
\end{IEEEbiography}

\end{document}